# The compressional beta effect:
# Analytical solution, numerical benchmark, and data analysis

Hing Ong[1]* and Paul E. Roundy[2]

[1]Department of Land, Air and Water Resources, University of California, Davis, Davis, California.

[2]Department of Atmospheric and Environmental Sciences, University at Albany, State University of New York, Albany, New York.

Corresponding author: Hing Ong (hxong@ucdavis.edu)

**Abstract**

This study derives a complete set of equatorially confined wave solutions from an anelastic equation set with the complete Coriolis terms, which include both the vertical and meridional planetary vorticity. The propagation mechanism can change with the effective static stability. When the effective static stability reduces to neutral, buoyancy ceases, but the role of buoyancy as an eastward-propagation mechanism is replaced by the compressional beta effect, i.e., vertical density-weighted advection of the meridional planetary vorticity. For example, the Kelvin mode becomes a compressional Rossby mode. Compressional Rossby waves are meridional vorticity disturbances that propagate eastward owing to the compressional beta effect. The compressional Rossby wave solutions can serve as a benchmark to validate the implementation of the nontraditional Coriolis terms (NCTs) in numerical models; with an effectively neutral condition and initial large-scale disturbances given a half vertical wavelength spanning the troposphere on Earth, compressional Rossby waves are expected to propagate eastward at a phase speed of 0.24 m s$^{-1}$. The phase speed increases with the planetary rotation rate and the vertical wavelength and also changes with the density scale height. Besides, the compressional beta effect and the meridional vorticity tendency are reconstructed using reanalysis data and regressed upon tropical precipitation filtered for the Madden–Julian oscillation (MJO). The results suggest that the compressional beta effect contributes 10.8% of the meridional vorticity tendency associated with the MJO in terms of the ratio of the minimum values.

## 1. Introduction

Theories about equatorially confined waves substantially explain the observed tropical large-scale variability of cloudiness and precipitation (Kiladis et al. 2009). Matsuno (1966) derived a set of equatorially confined wave solutions from the shallow water equation set. Silva Dias et al. (1983) derived a vertical normal mode transform through which the hydrostatic primitive equation set projects completely onto the shallow water equation set given rigid upper and lower boundaries. Although a rigid upper boundary does not exist, equatorially confined wave solutions derived from the hydrostatic primitive equation set (Holton and Hakim 2013) are equivalent to Matsuno's (1966) solutions assuming the rigid boundaries (Kiladis et al. 2009). The vertical normal mode transform (Silva Dias et al. 1983) established a theoretical foundation for applying Matsuno's (1966) model to tropical tropospheric large-scale flow. Wheeler and Kiladis (1999) demonstrated that large parts of the space-time spectra of the cloudiness variability conform to the dispersion relations of Matsuno (1966). Kiladis et al. (2009) summarized these theories and emphasized the concept of effective static stability felt by the waves. The effects of static stability as a source of restoring force on waves can be reduced when, in terms of anomalies associated with waves, diabatic heating or cooling due to increased or decreased moisture condensation partially offsets adiabatic cooling or warming due to upward or downward motion (Haertel and Kiladis 2004). Maher et al. (2019) suggested that Matsuno's model and the weak temperature gradient (WTG) model (e.g., Bretherton and Sobel 2003; Sobel et al. 2001; Yano and Bonazzola 2009) are two of the useful model hierarchies for understanding tropical atmospheric processes. These two hierarchies simplify the thermodynamics using different assumptions. In terms of convective coupling, Matsuno's model assumes that the vertical motion constrains the diabatic effects so that the static stability is effectively reduced, and the WTG model assumes that the diabatic effects force the vertical motion to the extent that the buoyancy ceases. Each of the hierarchies cannot be deduced to its complete form from each other. However, for Matsuno's model, reducing the effective static stability to neutral yields no buoyancy, so the model reaches the WTG balance but does not necessarily conform to the WTG model. Such an apparent intersection of the hierarchies motivates us to explore the effectively neutral condition.

The equatorially confined wave theory is based on an unforced framework. Though diabatic heating and cooling are involved, they are theoretically symmetric about the mean state and affect only the effective buoyancy frequency. In time scales of intraseasonal or longer, atmospheric flow is prone to dissipation, and a forced-dissipative framework is likely more analogous to most flows; for example, Gill's (1980) model simulates large-scale flow forced by diabatic heating. In such time scales, though unforced frameworks like Matsuno's (1966) cannot be excluded as a possible analog for the upper tropospheric flow (Roundy 2012; 2020), forced-dissipative frameworks like Gill's (1980) have been useful in understanding large-scale flow associated with the Madden–Julian oscillation (MJO, e.g., Adames and Kim 2016; Hayashi and Itoh 2012), the El Niño–Southern Oscillation (ENSO, e.g., Neelin et al. 1998), and the intertropical convergence zone (ITCZ, e.g., Ong and Roundy 2019; Vallis 2017).

Most of the forced-dissipative models assume the hydrostatic approximation following Gill (1980). The hydrostatic primitive equation set omits the nontraditional Coriolis terms (NCTs), which are terms involving the meridional planetary vorticity, $2\Omega \cos\vartheta$ ($\Omega$ and $\vartheta$ denote planetary rotation rate and latitude). NCTs are negligible when the buoyancy frequency is far larger than the meridional planetary vorticity (e.g., Müller 1989), which would be valid on Earth if the atmosphere were dry. However, later studies suggested that the buoyancy frequency can be effectively reduced by moist convection (e.g., Haertel and Kiladis 2004), and the validity of the omission of NCTs was reassessed by Hayashi and Itoh (2012) and Ong and Roundy (2019). These studies switched NCTs on and off in a linearized forced-dissipative model to simulate large-scale flow forced by a prescribed eastward-moving intraseasonal-oscillating heat source along the equator (Hayashi and Itoh 2012) and a prescribed zonally symmetric steady heat source (Ong and Roundy 2019). The results suggested that NCTs contribute 10% or more of the forced vertical vorticity fields through tilting the meridional planetary vorticity to the vertical. Moreover, Ong and Roundy (2020) accounted for the vertical NCT to correct the hypsometric equation, and the correction contributes ~ 5% of the tropical large-scale geopotential height variability. The effective buoyancy frequency is more difficult to estimate than length and depth scales. Thus, using the ratio of the NCT to the traditional Coriolis term in the zonal momentum equation as a measure to validate the hydrostatic approximation for large-scale flow, Ong and Roundy (2019) proposed a nondimensional parameter, $\hat{O} \equiv aD/\bar{Y}\bar{L}$, where the characteristic scaling variables for a heat source or sink are defined as follows: $a$, distance from planet center; $\bar{Y}$, distance of the corresponding subtropical jet from equator; $D$, vertical depth; and $\bar{L}$, meridional length. The hydrostatic approximation is valid only if $\hat{O}$ is small so that NCTs are negligible. Yet how do NCTs affect unforced equatorial waves? Also, can $\hat{O}$ measure the significance of NCTs in unforced equatorial waves?

Research about effects of NCTs on wave propagation began with a focus on the interior of stars and giant planets, and the following two important effects have been identified: topographic beta effect (e.g., Busse 1994; Gerkema et al. 2008; Heimpel et al. 2005; Yano 1998) and compressional beta effect (e.g., Gilman and Glatzmaier 1981; Glatzmaier et al. 2009; Verhoeven and Stellmach 2014). Considering vortex tubes parallel to the rotation axis spanning the interior confined by, typically, a spherical outer boundary, the topographic beta effect refers to vortex stretching due to radial motion. Busse's linear model (e.g., Busse 1994) is classical but oversimplifies the topographic beta effect (Yano 1998), and later studies (e.g., Heimpel et al. 2005) used numerical models to simulate this effect. On the other hand, considering local meridional vorticity, the compressional beta effect refers to vertical density-weighted advection of the meridional planetary vorticity. To illustrate, consider a positive meridional vorticity disturbance. To the east of the positive disturbance, in terms of the meridional planetary vorticity divided by density, the downward motion yields positive advection. Multiplying density converts this advection to increasing meridional relative vorticity via compression. The opposite occurs to the west. Consequently, the compressional beta effect transmits the meridional vorticity disturbance to the east. Focusing on the interior dynamics of giant planets, Glatzmaier et al. (2009) argued the importance of the compressional beta effect, which was coupled to the topographic beta effect using their numerical model. Using an unbounded linear model, Verhoeven and Stellmach (2014) untangled the compressional beta effect from coupling with the topographic beta effect. They referred to Rossby waves as driven by density-weighted advection of planetary vorticity in general. However, Rossby waves conventionally refer to waves driven by meridional advection of vertical planetary vorticity (e.g., Holton and Hakim 2013; Vallis 2017). Abiding by this convention, this paper refers to waves driven by the compressional beta effect as compressional Rossby waves. Verhoeven and Stellmach (2014) attempted to derive the dispersion relation of compressional Rossby waves. They found that the compressional beta effect transmits zonal vertical circulation to the east. However, their derivation is dynamically inconsistent (see Section 3) and is limited to a zonal vertical plane.

**Table 1.** Categories of equatorially confined wave solutions

|  | Hydrostatic | Quasi-hydrostatic | Fully nonhydrostatic |
|---|---|---|---|
| Shallow water | Matsuno (1966) | | |
| Boussinesq | | Fruman (2009) | Roundy and Janiga (2012) |
| Anelastic | Holton and Hakim (2013) | | The present study |

Research about effects of NCTs on the complete set of equatorially confined wave solutions has been in progress (Fruman 2009; Roundy and Janiga 2012). Fruman (2009) used a Boussinesq equation set including NCTs but not vertical acceleration (quasi-hydrostatic), and Roundy and Janiga (2012) further included vertical acceleration (fully nonhydrostatic). These two cases are similar for low frequency and long zonal wavelength. Categories of equatorially confined wave solutions are depicted in Table 1. In the Boussinesq models, NCTs widen the meridional decay length scale of the equatorially confined waves. At a certain longitude, NCTs tilt the lines of constant phase upward and poleward, so the wave phases propagate either equatorward and upward or poleward and downward, while the meridional wave energy propagation is zero. However, NCTs do not affect the dispersion relations of any subset of the equatorially confined wave solutions in the Boussinesq models (Fruman 2009; Roundy and Janiga 2012). The reason may be that the meridional planetary vorticity divided by density is constant in the Boussinesq models, and a gradient of the meridional planetary vorticity divided by density is necessary for the compressional beta effect (e.g., Gilman and Glatzmaier 1981; Glatzmaier et al. 2009; Verhoeven and Stellmach 2014) to change the dispersion relations. Previous studies about effects of NCTs on waves on an f-plane (Kasahara 2003; Kohma and Sato 2013) are also useful for this study; especially, Kohma and Sato (2013) used an anelastic equation set. The solutions on a beta plane should reduce to the solutions on an f-plane when $\beta \to 0$.

Development of dynamical cores for atmospheric models usually benefits from research about deterministic initial value problems. For example, numerical benchmarks of baroclinic waves (e.g., Jablonowski and Williamson 2006; Ullrich et al. 2014) are widely used to test the model performance in the midlatitudes. On the other hand, in the tropics, simply testing the dry dynamics over-stratifies the atmosphere, but adding full moist processes overcomplicates the benchmarking test. This conundrum motivates Reed and Jablonowski (2012) to design simplified moist physical parameterization for testing the tropical performance. To further eliminate physical parameterization, this study tunes the dynamical parameters to make the dry dynamical core more relevant to the moist tropical atmosphere. Research about analytical wave solutions emerging from the compressional beta effect can be applied to validate the implementation of NCTs in the dynamical cores of atmospheric models. Such research can be important because many model developers are restoring NCTs, including DWD's ICOsahedral Non-hydrostatic model (ICON, Borchert et al. 2019), GFDL's Finite-Volume Cubed-Sphere Dynamical Core (FV3, Hann-Ming Henry Juang 2019, personal communication), and NCAR's Model for Prediction Across Scales (MPAS, William C. Skamarock 2019, personal communication). Borchert et al. (2019) applied a numerical benchmark of baroclinic waves (Ullrich et al. 2014) and an analytical benchmark of acoustic waves. This study attempts to propose a more useful benchmark featuring exact wave solutions that can only exist with NCTs and dynamical parameters that eliminate buoyancy.

This paper is organized as follows. Section 2 discusses an anelastic equation set used in the following sections. Section 3 derives the compressional Rossby wave solution. Section 4 derives the complete set of equatorially confined wave solutions. Section 5 applies the compressional Rossby wave solution to design a benchmarking test and presents results using the MPAS. Section 6 demonstrates how to analyze the compressional beta effect from data by exploring its contribution to meridional vorticity tendency associated with the MJO. Section 7 presents summary and discussion.

## 2. Anelastic Equation Set

An anelastic equation set formulated in Lipps and Hemler (1982) is used because vorticity dynamics govern this dynamical system (Jung and Arakawa 2008). Linearize the equation set around a motionless stratified reference state with the complete Coriolis terms on the equatorial beta plane, where $2\Omega \cos \vartheta$ reduces to $2\Omega$ while $2\Omega \sin \vartheta$ reduces to $\beta y$; $\beta = 2\Omega/a$;

$$\frac{\partial b}{\partial t} + \widetilde{N}^2 w = 0, \tag{1a}$$

$$\frac{\partial u}{\partial t} - \beta y v + 2\Omega w + \frac{\partial \varphi}{\partial x} = 0, \tag{1b}$$

$$\frac{\partial v}{\partial t} + \beta y u + \frac{\partial \varphi}{\partial y} = 0, \tag{1c}$$

$$\epsilon \frac{\partial w}{\partial t} - 2\Omega u + \frac{\partial \varphi}{\partial z} - b = 0, \tag{1d}$$

$$\frac{\partial u}{\partial x} + \frac{\partial v}{\partial y} + \frac{\partial w}{\partial z} - \frac{w}{H} = 0. \tag{1e}$$

The variables are defined as follows: $u$, zonal velocity; $v$, meridional velocity; $w$, vertical velocity; $b$, buoyancy; and $\varphi$, potential-temperature-weighted perturbation Exner function (a pressure-like perturbation proposed by Lipps and Hemler 1982). The coordinates are geometric where $z$ denotes geopotential height. The parameters are defined as follows: $N$, buoyancy frequency; $1/H \equiv -\mathrm{d}\ln\rho/\mathrm{d}z$, inverse scale height of reference density, $\rho$. To validate the equatorial beta plane approximation, $a$ (distance from planet center, used to define $\beta$, $x$, $y$, and $z$) must be larger than the characteristic meridional width and vertical depth. There is neither forcing nor dissipation in equations (1), but given $\widetilde{N} \equiv \sqrt{\alpha}N$, there can be diabatic heating and cooling depending on $\alpha$, which is a nondimensional effective buoyancy parameter. $\alpha = 1$ sets vertical motion dry-adiabatic, and $\alpha \in [0,1)$ reduces the effect of vertical motion on buoyancy; $\alpha = 0$ is the neutral limit. Parameter $\widetilde{N}$ is defined as the effective buoyancy frequency. $\epsilon$ is a nondimensional vertical acceleration parameter. $\epsilon = 1$ and $0$ set the dynamical system fully nonhydrostatic and quasi-hydrostatic. $\epsilon$ serves as a dynamical tracer for the vertical acceleration term during the derivation. Terms with explicit $\Omega$ and $\beta$ are the nontraditional and traditional Coriolis terms.

The energy equation is derived because this study emphasizes energy constraints including energy conservation during wave propagation and energy confinement in the equatorial region. Apply equation (1e) to the sum of the following: (1a) × $\rho b/\widetilde{N}^2$ + (1b) × $\rho u$ + (1c) × $\rho v$ + (1d) × $\rho w$, and average over a wave period (overbar);

$$\frac{\partial}{\partial t}\left[\frac{\rho}{2}\left(\frac{b^2}{\widetilde{N}^2} + u^2 + v^2 + \epsilon w^2\right)\right] + \frac{\partial}{\partial x}(\overline{\rho\varphi u}) + \frac{\partial}{\partial y}(\overline{\rho\varphi v}) + \frac{\partial}{\partial z}(\overline{\rho\varphi w}) = 0. \tag{2}$$

Equation (2) states a form of local energy conservation; local tendency of total energy, $\frac{\rho}{2}\left(\frac{b^2}{\widetilde{N}^2} + u^2 + v^2 + \epsilon w^2\right)$, equals to three-dimensional convergence of energy flux, $\rho\varphi u$, $\rho\varphi v$, and $\rho\varphi w$ for zonal, meridional, and vertical. With periodic and radiation boundary conditions in zonal and vertical directions, to conserve energy during zonal vertical wave propagation, total energy and zonal vertical energy flux must be constant at a certain latitude for every single plane wave solution. Accordingly, the amplitude of $u$, $v$, $w$, $b$, and $\varphi$ must *increase exponentially with altitude* to be inversely proportional to the square root of $\rho$ for every single plane wave solution. To confine energy in an unbounded equatorial region, for any combinations of wave solutions, total energy must decay to zero as $y \to \pm\infty$, and meridional energy flux must be zero. Consequently, the phases of $\varphi$ and $v$ must be *in quadrature* so that their inner product is zero.

The meridional vorticity equation is also derived because it simplifies the derivation of compressional Rossby wave solutions. Apply equation (1e) to the following: $\partial$ (1b) / $\partial z - \partial$ (1d) / $\partial x$;

$$\frac{\partial}{\partial t}\left(\frac{\partial u}{\partial z} - \epsilon\frac{\partial w}{\partial x}\right) + 2\Omega\frac{w}{H} - 2\Omega\frac{\partial v}{\partial y} - \beta y\frac{\partial v}{\partial z} + \frac{\partial b}{\partial x} = 0. \tag{3}$$

Equation (3) states that meridional relative vorticity, $\frac{\partial u}{\partial z} - \epsilon\frac{\partial w}{\partial x}$, changes in time in response to the following mechanisms; $-2\Omega\frac{w}{H}$, vertical density-weighted advection of meridional planetary vorticity, i.e., the compressional beta effect; $2\Omega \frac{\partial v}{\partial y}$, meridional stretching of meridional planetary vorticity; $\beta y \frac{\partial v}{\partial z}$, tilting of planetary vorticity from vertical to meridional; $-\frac{\partial b}{\partial x}$, buoyancy generation. To gain more insight into the compressional beta effect, rewrite the term; $-2\Omega \frac{w}{H} = 2\Omega w \frac{\mathrm{d}\ln\rho}{\mathrm{d}z} = -\rho w \frac{\mathrm{d}}{\mathrm{d}z}\left(\frac{2\Omega}{\rho}\right)$. In this form, the vertical advection operator, $-w\frac{\mathrm{d}}{\mathrm{d}z}$, multiplies density, and the advected quantity is the meridional planetary vorticity divided by density.

### 3. Compressional Rossby Waves

To derive compressional Rossby waves, ignore terms involving $v$ and $b$ in equation (3). This step isolates the compressional beta effect from the complex equation set, which is the subject of Section 4. Ignoring $\partial v/\partial y$ enables rewriting equation (3) in terms of zonal vertical mass stream function, $\Psi$, where $\rho u \equiv \partial \Psi/\partial z$ and $\rho w \equiv -\partial \Psi/\partial x$;

$$\frac{\partial}{\partial t}\left(\epsilon \frac{\partial^2 \Psi}{\partial x^2} + \frac{\partial^2 \Psi}{\partial z^2} + \frac{1}{H}\frac{\partial \Psi}{\partial z}\right) - \frac{2\Omega}{H}\frac{\partial \Psi}{\partial x} = 0, \tag{4}$$

where $\frac{1}{H}\frac{\partial \Psi}{\partial z}$ can be interpreted as a compressional effect on the stream function because it emerges from the reference density variations.

Assume zonal vertical plane wave solutions to equation (4); $\Psi = \widehat{\Psi} \exp(-z/2H) \exp[i(kx + mz - \omega t)]$. The factor $\exp(-z/2H)$ ensures energy conservation during vertical propagation because $\rho$ and the amplitude of $w$ have factors of $\exp(-z/H)$ and $\exp(z/2H)$. Plug the assumed solutions into equation (4), and rearrange;

$$\frac{\omega}{k} = \frac{2\Omega}{H}\left(\epsilon k^2 + m^2 + \frac{1}{4H^2}\right)^{-1}. \tag{5}$$

Equation (5) states the dispersion relation of compressional Rossby waves. The phase speed ($\omega/k$) is eastward and increases with the planetary rotation rate ($\Omega$), the vertical wavelength ($2\pi/m$), and the zonal wavelength ($2\pi/k$); $k$ is insignificant for large-scale flow. The zonal phase speed also changes with the density scale height ($H$), yet not monotonically; for $m^2 > 1/4H^2$, the zonal phase speed increases with decreasing $H$, and vice versa. For large-scale compressional Rossby waves on Earth with a half vertical wavelength spanning an effectively neutral troposphere, the zonal phase speed is 0.24 m s$^{-1}$, given $\Omega = 7.292 \times 10^{-5}$ s$^{-1}$, $H = 9.1$ km, and $2\pi/m = 25$ km. Superposing incident and reflected waves against a rigid lower boundary, the solution becomes:

$$w = w_0 \exp\left(\frac{z}{2H}\right) \sin(mz) \sin(kx - \omega t), \tag{6a}$$

$$\varphi = \varphi_0 \exp\left(\frac{z}{2H}\right) \cos\left(mz + \arctan\frac{\frac{\omega}{2H} - 2\Omega k}{m\omega}\right) \cos(kx - \omega t), \tag{6b}$$

$$u = u_0 \exp\left(\frac{z}{2H}\right) \cos\left(mz + \arctan\frac{\frac{\omega}{2H} - 2\Omega k}{m\omega} + \arctan\frac{2\Omega m}{\frac{\Omega}{H} - \epsilon\omega k}\right) \cos(kx - \omega t), \tag{6c}$$

$$w_0 = \frac{k}{\sqrt{4\Omega^2 - \epsilon\omega^2}} \varphi_0, \tag{6d}$$

$$u_0 = \frac{\sqrt{\frac{\Omega^2}{H^2} + \epsilon\omega^2 k^2 - \epsilon\frac{2\Omega\omega k}{H} + 4\Omega^2 m^2}}{4\Omega^2 - \epsilon\omega^2} \varphi_0, \tag{6e}$$

where $\varphi_0$, $w_0$, and $u_0$ denote amplitudes of $\varphi$, $w$, and $u$. Figure 1 shows snapshots of the analytical solution of the zonal vertical structures of such waves. In Figure 1a, the downward motion yields positive density-weighted advection of the meridional planetary vorticity, and the upward motion yields the opposite. Hence, the meridional vorticity disturbances propagate eastward. The dispersion relation derived by Verhoeven and Stellmach (2014) resembles equation (5) but lacks the term $1/4H^2$ because they ignored the compressional effect on the stream function while considering the compressional beta effect; hence, their derivation is dynamically inconsistent. Verhoeven and Stellmach (2014) mentioned one of the restrictions on the validity of their solution; $m^2 \gg \frac{1}{4H^2}$. Yet even if $m^2 \gg \frac{1}{4H^2}$, their solution does not conserve energy when the waves propagate vertically by a distance of order $H$. If $m^2 \leq \frac{1}{4H^2}$, their solution will have a remarkable fast bias in terms of the phase speed.

In equations (6b) and (6c), the vertical phase of $u$ is shifted from the vertical phase of $\varphi$ by $\arctan \frac{2\Omega m}{\frac{\Omega}{H} - \epsilon \omega k}$. In Figure 1b, a low-$\varphi$ region is located above a low-$u$ region and below a high-$u$ region, and vice versa. This relation is consistent with Ong and Roundy (2020), who introduced NCTs to the hypsometric equation and showed that easterly winds in a layer correspond to low pressure perturbations above the layer or high below it. The structure in Figure 1b is a signature of compressional Rossby waves, which is different from Kelvin waves, where $u$ and $\varphi$ are in phase (Figure 1c).

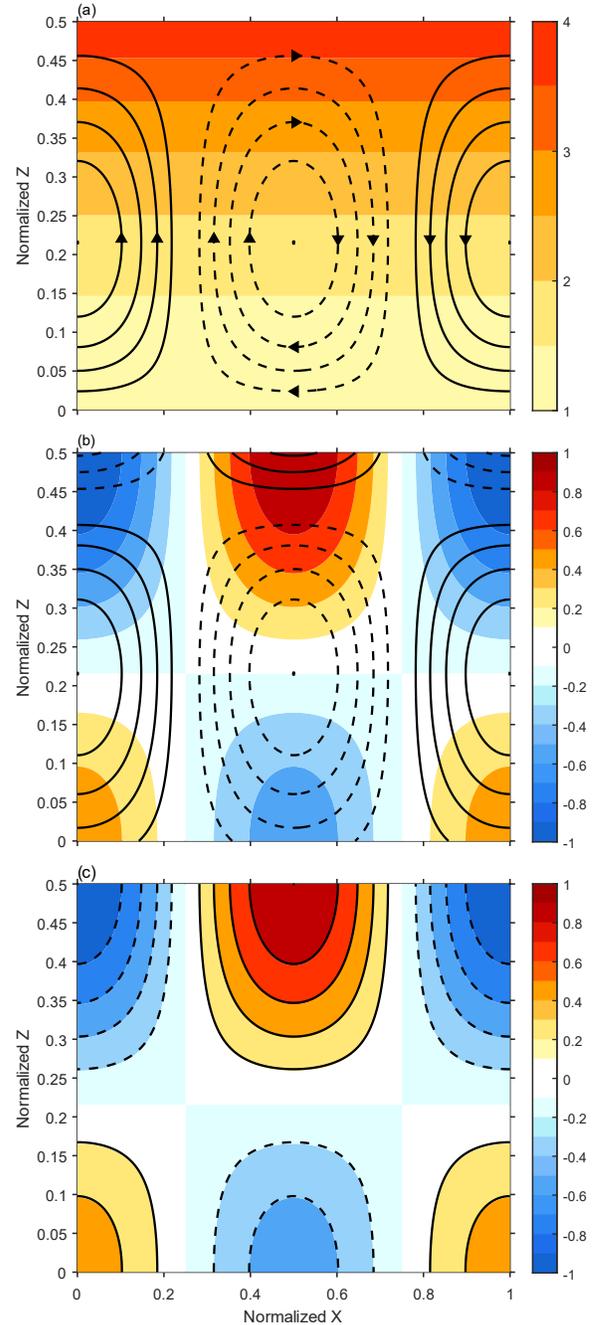

**Figure 1**. Snapshots of the zonal vertical structures of the analytical solution of (a-b) the compressional Rossby waves ($\widetilde{N} = 0$) and (c) the Kelvin waves ($\widetilde{N} = N$). In panel (a), the contours denote the mass stream function, and the arrows denote the mass flux direction. The shading denotes the meridional planetary vorticity divided by density normalized by the surface value. In panels (b-c), the contours denote $\varphi$ (a pressure-like perturbation), and the shading denotes the zonal wind. The dashed contours denote negative values (negative stream function corresponds to positive meridional relative vorticity), and the zero contours are omitted. The length and depth scales are normalized by the wavelengths.

## 4. Complete Set of Equatorially Confined Waves

To derive the complete set of equatorially confined waves, assume zonal vertical plane wave solutions to equation (1) that vary meridionally; $\{u, v, w, b, \varphi\} = \{\hat{u}(y), \hat{v}(y), \hat{w}(y), \hat{b}(y), \hat{\varphi}(y)\} \exp(z/2H) \exp[i(kx + mz - \omega t)]$. The amplitudes vary vertically and meridionally. Vertically, the factor $\exp(z/2H)$ ensures energy conservation. Meridionally, the hatted factors are unknown and will be solved given the necessary conditions for energy confinement in the equatorial region. Plug the assumed solutions into equation (1);

$$i\omega \hat{b} = \tilde{N}^2 \hat{w}, \tag{7a}$$

$$i\omega \hat{u} = -\beta y \hat{v} + 2\Omega \hat{w} + ik\hat{\varphi}, \tag{7b}$$

$$-i\omega \hat{v} + \beta y \hat{u} + \frac{d\hat{\varphi}}{dy} = 0, \tag{7c}$$

$$-i\omega \epsilon \hat{w} - 2\Omega \hat{u} + \left(\frac{1}{2H} + im\right)\hat{\varphi} - \hat{b} = 0, \tag{7d}$$

$$ik\hat{u} + \frac{d\hat{v}}{dy} + \left(-\frac{1}{2H} + im\right)\hat{w} = 0. \tag{7e}$$

Because the relation between $\hat{\varphi}$ and $\hat{v}$ is the pivot to determine the necessary conditions for the energy confinement, $\hat{b}$, $\hat{u}$, and $\hat{w}$ are eliminated through the following steps in order: multiply equations (7c-e) by $i\omega$, plug equations (7a-b) in to eliminate $\hat{b}$ and $\hat{u}$, multiply the new equations (7c) and (7e) by $(\epsilon\omega^2 - \tilde{N}^2 - 4\Omega^2)$, and plug the new equation (7d) in to eliminate $\hat{w}$;

$$[\omega^2(\epsilon\omega^2 - \tilde{N}^2 - 4\Omega^2) + \beta^2 y^2(\tilde{N}^2 - \epsilon\omega^2)]\hat{v} + \left[2\Omega\beta y m\omega + ik\beta y\left(\epsilon\omega^2 - \tilde{N}^2 - \frac{\Omega\omega}{Hk}\right)\right]\hat{\varphi}$$
$$+ i\omega(\epsilon\omega^2 - \tilde{N}^2 - 4\Omega^2)\frac{d\hat{\varphi}}{dy} = 0, \tag{8a}$$

$$\left[-k^2(\tilde{N}^2 - \epsilon\omega^2) - k\frac{2\Omega\omega}{H} + \omega^2\left(m^2 + \frac{1}{4H^2}\right)\right]\hat{\varphi} = \left[2\Omega\beta y m\omega - ik\beta y\left(\epsilon\omega^2 - \tilde{N}^2 - \frac{\Omega\omega}{Hk}\right)\right]\hat{v}$$
$$+ i\omega(\epsilon\omega^2 - \tilde{N}^2 - 4\Omega^2)\frac{d\hat{v}}{dy}. \tag{8b}$$

Given any $y$ that is real, according to equation (8a), $\hat{\varphi} = 0$ yields trivial solutions because $\hat{v} = 0$ must be true. According to equation (8b), $\hat{\varphi} \neq 0$ yields two types of nontrivial solutions, zero-$\hat{v}$ and nonzero-$\hat{v}$. Also, the zero-$\hat{v}$ and nonzero-$\hat{v}$ cases require zero-$K$ and nonzero-$K$, where $K \equiv -k^2(\tilde{N}^2 - \epsilon\omega^2) - k\frac{2\Omega\omega}{H} + \omega^2\left(m^2 + \frac{1}{4H^2}\right)$. Subsections 4a and 4b solve these two cases separately, and Subsection 4c discusses the solutions.

### a. Zero-$\hat{v}$ Case

Apply $\hat{v} = 0$ to equation (8);

$$\left[2\Omega\beta y m\omega + ik\beta y\left(\epsilon\omega^2 - \tilde{N}^2 - \frac{\Omega\omega}{Hk}\right)\right]\hat{\varphi} + i\omega(\epsilon\omega^2 - \tilde{N}^2 - 4\Omega^2)\frac{d\hat{\varphi}}{dy} = 0, \tag{9a}$$

$$\left[-k^2(\tilde{N}^2 - \epsilon\omega^2) - k\frac{2\Omega\omega}{H} + \omega^2\left(m^2 + \frac{1}{4H^2}\right)\right]\hat{\varphi} = 0. \tag{9b}$$

Integrating equation (9a) yields the zero-$\hat{v}$ solution for $\hat{\varphi}$, and plugging this into the original assumed solution yields the following:

$$\varphi = \varphi_0 \exp\left(\frac{z}{2H} - \frac{\tilde{N}^2 + \frac{\Omega\omega}{Hk} - \epsilon\omega^2}{\tilde{N}^2 + 4\Omega^2 - \epsilon\omega^2} \frac{\beta k}{\omega} \frac{y^2}{2}\right) \exp\left[i\left(kx - \omega t + mz + \frac{-2\Omega\beta m}{\tilde{N}^2 + 4\Omega^2 - \epsilon\omega^2} \frac{y^2}{2}\right)\right]. \quad (10)$$

Equation (9b) yields the dispersion relation of the zero-$\hat{v}$ solution;

$$-k^2(\tilde{N}^2 - \epsilon\omega^2) - k\frac{2\Omega\omega}{H} + \omega^2\left(m^2 + \frac{1}{4H^2}\right) = 0. \quad (11)$$

Equations (10) and (11) are consistent with the Kelvin wave solutions in previous studies (Fruman 2009; Holton and Hakim 2013; Kohma and Sato 2013; Roundy and Janiga 2012) when certain limits are taken. At the hydrostatic limit, i.e., $\epsilon \to 0$ and $\Omega \to 0$, equations (10) and (11) reduce to the solutions of Holton and Hakim (2013). At the Boussinesq limit, i.e., $H \to \infty$, equations (10) and (11) reduce to the solutions of Roundy and Janiga (2012), which further reduce to the solutions of Fruman (2009) at the quasi-hydrostatic limit, i.e., $\epsilon \to 0$. Furthermore, equation (11) is equivalent to equation (33) of Kohma and Sato (2013), who suggested that these waves are not trapped by a zonal boundary at the equator using an f-plane. However, equations (10) and (11) suggest that these waves are trapped on the equatorial beta plane only if propagating eastward; given equation (11), $\frac{\tilde{N}^2 + \frac{\Omega\omega}{Hk} - \epsilon\omega^2}{\tilde{N}^2 + 4\Omega^2 - \epsilon\omega^2} \frac{\beta k}{\omega} = \frac{\beta\sqrt{(\tilde{N}^2 - \epsilon\omega^2)(m^2 + \frac{1}{4H^2}) + \frac{\Omega^2}{H^2}}}{\tilde{N}^2 + 4\Omega^2 - \epsilon\omega^2} > 0$ in equation (10) if and only if $\frac{\omega}{k} > 0$ and $\frac{\epsilon k^2}{m^2 + \frac{1}{4H^2}} < \sqrt{1 + \frac{\tilde{N}^2 H^2}{\Omega^2}\left(\epsilon k^2 + m^2 + \frac{1}{4H^2}\right)}$. The second condition only restrains a large aspect ratio ($\frac{\epsilon k^2}{m^2 + \frac{1}{4H^2}}$) from the equatorial confinement.

However, the zero-$\hat{v}$ waves are not Kelvin waves. To illustrate, at the neutral limit, i.e., $\tilde{N} \to 0$, equation (11) reduces to equation (5), i.e., compressional Rossby waves. Moreover, taking this limit for equation (10) suggest that the compressional Rossby waves are equatorially confined for an aspect ratio smaller than unity. With the effective static stability increasing from neutral, equation (11) approaches the canonical solution for Kelvin waves, with a continuum of hybrid forms in between. Kelvin wave dynamics dominate if the effective buoyancy frequency is larger than the meridional planetary vorticity. All zero-$\hat{v}$ waves with a small aspect ratio, i.e., $\epsilon k^2 \ll m^2 + \frac{1}{4H^2}$, are nondispersive in the zonal direction.

b. Nonzero-$\hat{v}$ Case

Derivations to be elaborated in this section show that the nonzero-$\hat{v}$ solutions of equation (8) can be decomposed as $\hat{v} \equiv v_0 V\left(\frac{y}{L}\right) \exp\left(\frac{-y^2}{2L^2}\right) \exp\left(\frac{i\Gamma y^2}{2}\right)$, where the four factors denote amplitude of $v$, meridional stationary oscillator, meridional decay function, and meridional propagation oscillator. $\Gamma$ can be interpreted as a meridional propagation parameter; phases propagate poleward for positive $\Gamma$, and vice versa. $\Gamma$ can also be interpreted as a meridional tilting parameter; lines of constant phase tilt upward and poleward if the signs of $\Gamma$ and $m$ are opposite, and vice versa. To discuss energy constraints on $\Gamma$, plug the decomposition and $K \equiv -k^2(\tilde{N}^2 - \epsilon\omega^2) - k\frac{2\Omega\omega}{H} + \omega^2\left(m^2 + \frac{1}{4H^2}\right)$ into equation (8b);

$$K\hat{\varphi} = [2\Omega\beta ym\omega - \Gamma y\omega(\epsilon\omega^2 - \tilde{N}^2 - 4\Omega^2) - ik\beta y\left(\epsilon\omega^2 - \tilde{N}^2 - \frac{\Omega\omega}{Hk}\right)$$

$$-i\frac{y}{L^2}\omega(\epsilon\omega^2 - \tilde{N}^2 - 4\Omega^2) + i\frac{1}{V}\frac{dV}{dy}\omega(\epsilon\omega^2 - \tilde{N}^2 - 4\Omega^2)]\hat{v}. \quad (12)$$

To prevent any meridional energy flux, if $\hat{\varphi}$ is real, $\hat{v}$ must be imaginary. To satisfy equation (12), if $\hat{\varphi}$ is real, the rhs of equation (12) must be real. Consequently, given $\hat{\varphi}$ is real without loss of generality (assuming any complex $\hat{\varphi}$ yields the same conclusion), on the rhs of equation (12), $-\Gamma y \omega (\epsilon \omega^2 - \tilde{N}^2 - 4\Omega^2)$ must cancel $2\Omega\beta y m \omega$, which constrains the meridional propagation (tilting) parameter;

$$\Gamma = \frac{-2\Omega\beta m}{\tilde{N}^2 + 4\Omega^2 - \epsilon \omega^2}. \tag{13}$$

Accordingly, the meridional phase propagation is *nonzero* as in equation (13) if and only if the meridional energy propagation is *zero*. Equation (13) is equivalent to equation (18) of Roundy and Janiga (2012); thus, the meridional phase propagation is independent from the reference density variations. Moreover, because $\tilde{N}^2 + 4\Omega^2 - \epsilon\omega^2 > 0$ for all real solutions, $\Gamma$ and $m$ are opposite signed. Consequently, Fruman's (2009) result of upward and poleward tilting of lines of constant phase also applies to the less-approximated case in the present study (Table 1).

To solve for $V$ and $L$, multiply equation (8a) by $K$, plug equation (12) into it, and rearrange;

$$(\tilde{N}^2 + 4\Omega^2 - \epsilon\omega^2)\frac{d^2}{dy^2}\left[V\exp\left(\frac{-y^2}{2L^2}\right)\right] + \left[\left(k^2 + \frac{k\beta}{\omega}\right)\left(\epsilon\omega^2 - \tilde{N}^2 - \frac{\Omega\omega}{Hk}\right) + \omega^2\left(m^2 + \frac{1}{4H^2} - \frac{\Omega k}{H\omega}\right) - \left(m^2 + \frac{1}{4H^2} - \frac{4\Omega^2 m^2}{\tilde{N}^2 + 4\Omega^2 - \epsilon\omega^2}\right)\beta^2 y^2\right]V\exp\left(\frac{-y^2}{2L^2}\right) = 0. \tag{14}$$

Then, to apply known solutions to equation (14), nondimensionalize it by plugging $y \equiv LY$ into it. This yields a form of Hermite's equation, $\frac{d^2V}{dY^2} - 2Y\frac{dV}{dY} + \lambda V = 0$, where

$$L^2 = \frac{\tilde{N}^2 + 4\Omega^2 - \epsilon\omega^2}{\beta\sqrt{(\tilde{N}^2 - \epsilon\omega^2)\left(m^2 + \frac{1}{4H^2}\right) + \frac{\Omega^2}{H^2}}}, \tag{15a}$$

$$\lambda = \frac{L^2}{\tilde{N}^2 + 4\Omega^2 - \epsilon\omega^2}\left[\left(k^2 + \frac{k\beta}{\omega}\right)\left(\epsilon\omega^2 - \tilde{N}^2 - \frac{\Omega\omega}{Hk}\right) + \omega^2\left(m^2 + \frac{1}{4H^2} - \frac{\Omega k}{H\omega}\right)\right] - 1. \tag{15b}$$

The solutions for $V$ are the physicists' Hermite polynomials, $H_n$, where $n = 0, 1, 2, \ldots$ (e.g., Vallis 2017). Plugging this into the original assumed solution yields the following:

$$v = v_0 H_n\left(\frac{y}{L}\right)\exp\left(\frac{z}{2H} - \frac{y^2}{2L^2}\right)\exp\left[i\left(kx - \omega t + mz + \frac{\Gamma y^2}{2}\right)\right]. \tag{16}$$

For each $n$, solutions exist if and only if $\lambda = 2n$, which yields the dispersion relations;

$$-\left(k^2 + \frac{k\beta}{\omega}\right)\left(\tilde{N}^2 + \frac{\Omega\omega}{Hk} - \epsilon\omega^2\right) + \omega^2\left(m^2 + \frac{1}{4H^2} - \frac{\Omega k}{H\omega}\right) = (2n+1)\frac{\tilde{N}^2 + 4\Omega^2 - \epsilon\omega^2}{L^2}. \tag{17}$$

Equations (13) and (15) through (17) are consistent with the non-Kelvin wave solutions in previous studies (Fruman 2009; Holton and Hakim 2013; Roundy and Janiga 2012) when certain limits are taken. A subset of the dispersion relations where $K = 0$ is discarded because the derivation of equation (14) requires $K \neq 0$. $L^2 > 0$ is true for all results discussed below.

### c. Discussion

The zonal temporal dispersion relations of the zero-$\hat{v}$ and nonzero-$\hat{v}$ cases are depicted together in Figure 2, given $\Omega = 7.292 \times 10^{-5}$ s$^{-1}$, $H = 9.1$ km, and $2\pi/m = 25$ km. In the strongly stable case (Figure 2a), all the modes appear like Matsuno's (1966) modes with an equivalent depth of 33 m, and the inclusion of NCTs does not make a noticeable difference in terms of the dispersion relations and the spatial structure. Such an equivalent depth lies within the canonical convectively coupled equatorial wave bands on Earth (e.g., Wheeler and Kiladis 1999). In the neutral case (Figure 2b), the zero-$\hat{v}$ and nonzero-$\hat{v}$ modes appear like the Kelvin and Yanai ($n = 0$, mixed Rossby-gravity) modes in Figure 2a, but the compressional beta effect replaces buoyancy as the eastward-propagation mechanism. Also, in Figure 2b, the westward inertio-gravity (high wavenumber and high frequency) modes in Figure 2a disappear because buoyancy is zero but is a fundamental restoring force of these waves. Moreover, in Figure 2b, the Rossby ($n > 0$ and low frequency) modes in Figure 2a coincide $K = 0$ so are discarded. For the zero-$\hat{v}$ mode (Figure 2c), with decreasing $\widetilde{N}$, the zonal phase speed decreases linearly without NCTs but nonlinearly with NCTs; in the latter case, the decreasing rate of phase speed decreases so that the phase speed approaches 0.24 m s$^{-1}$ instead of zero. For all the modes transitioning from Figure 2a to 2b, see the animation in mp4 format in the supplemental material, where black and red curves denote dispersion relations with and without NCTs. Except the last frame of the animation (Figure 2b), sound of piano is played at a sound frequency proportional to the effective buoyancy frequency used to plot every frame. With decreasing $\widetilde{N}$, the zonal phase speed of all modes decreases, and the dispersion curves with and without NCTs separate farther. Overall, the contributions of NCTs become noticeable when the effective buoyancy frequency becomes comparable or smaller than the meridional planetary vorticity, which is consistent with Müller (1989).

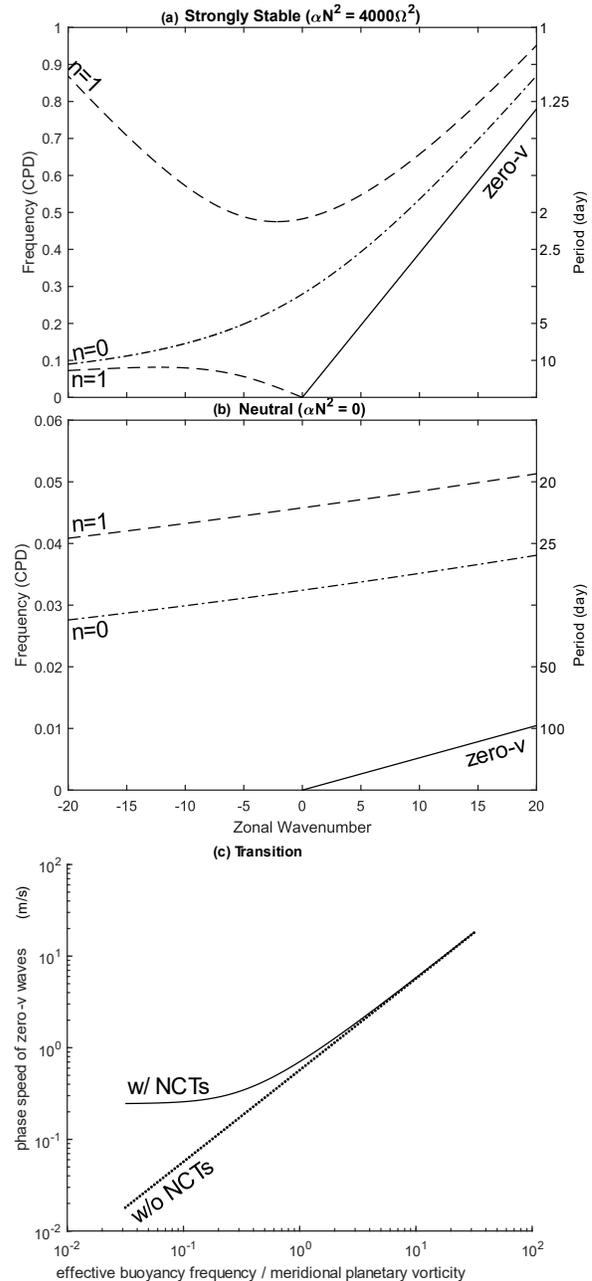

**Figure 2**. Zonal temporal dispersion relations of the equatorially confined wave solutions for (a) a strongly stable case and (b) the neutral case. Panel (c) depicts the transition of the zonal phase speed of the zero-$\hat{v}$ waves with and without NCTs from slightly stable to strongly stable.

The results suggest that $\hat{O}$ number (Ong and Roundy 2019) can measure the significance of NCTs in unforced equatorial waves. To estimate $\hat{O}$ number, choose $L$ as the characteristic $\bar{Y}$ and $\bar{L}$, and $2H$ as the characteristic $D$. Then, plug these choices and equation (15a) into $\hat{O} \equiv aD/\bar{Y}\bar{L}$, and assume low frequency where $\omega^2 \ll 4\Omega^2$. For the neutral case, $\hat{O} = 1$; in words, NCTs are on the leading order. For a strongly stable case where $2\Omega/\tilde{N} \to 0$, $\hat{O} \sim 2\Omega/\tilde{N}$; in words, NCTs are negligible, so Matsuno's (1966) solutions, using the hydrostatic approximation, can become valid.

## 5. Benchmarking Test

To test the model performance with the implementation of NCTs under a neutral condition, we choose the compressional Rossby wave solutions in Section 3 as a benchmark because the model configuration is simpler than the solutions in Section 4. The spatial domain is a zonal vertical rectangle. The lateral boundaries are periodic, and the upper and lower boundaries are rigid. The planetary vorticity has a northward component but no vertical component, i.e., using the generalized equatorial f-plane. We make the planetary rotation rate much faster to save process time; the wave period becomes as short as 86,400 s. The basic state is hydrostatic and motionless. The initial perturbations are set using equations (5) and (6). Table 2 lists the parameters for the benchmarking test.

**Table 2.** Parameters used in the benchmarking test

| | |
|---|---|
| $\Omega$ (planetary rotation rate) | $6.973339 \times 10^{-3}$ s$^{-1}$ |
| $g$ (gravity acceleration) | 9.80616 m s$^{-2}$ |
| $R$ (gas constant for dry air) | 287.0 J kg$^{-1}$ K$^{-1}$ |
| $T$ (basic-state temperature) | 311.0 K |
| $H$ (density scale height) | $RT/g \cong 9.1 \times 10^3$ m |
| $\kappa$ (Poisson constant) | 0 |
| $p_b$ (basic-state pressure at the bottom) | $1.0 \times 10^5$ Pa |
| $L_x$ (domain width) | $2.0 \times 10^6$ m |
| $k$ (zonal wavenumber) | $2\pi/L_x$ |
| $L_z$ (domain depth) | 12,721 m (fully compressible) <br> 12,500 m (anelastic) |
| $m$ (vertical wavenumber) | $\pi/L_z$ |
| $u_0$ (initial perturbation amplitude of zonal velocity) | 0.09 m s$^{-1}$ |

For the thermodynamics, we aim to eliminate buoyancy. A possible way is to initiate the test with constant potential temperature, but this drastically enhances the vertical decrease of the density scale height. Instead, we use a barotropic ideal gas whose thermodynamic properties fit our goal; its heat capacity is infinity, so an isothermal atmosphere becomes isentropic because its Poisson constant is zero. $\varphi$ for such a gas denotes perturbation of pressure divided by basic-state density. For a fully compressible model, its speed of sound is $\sqrt{gH}$, where $g$ denotes gravity acceleration, and equation (1e) becomes:

$$\frac{1}{gH}\frac{\partial \varphi}{\partial t} + \frac{\partial u}{\partial x} + \frac{\partial w}{\partial z} - \frac{w}{H} = 0. \tag{18}$$

While the structures in equation (6) still apply, the dispersion relation becomes:

$$\frac{\omega}{k} = \frac{2\Omega}{H}\left(\epsilon k^2 + m^2 + \frac{1}{4H^2} + \frac{4\Omega^2}{gH} - \epsilon\frac{\omega^2}{gH}\right)^{-1}. \tag{19}$$

Compressional Rossby waves propagate slightly slower in the fully compressible case as equation (19) than the anelastic case as equation (5). In Table 2, different values of $m$ are given for the two cases so that the wave period remains 86,400 s. In practice, $-\epsilon \frac{\omega^2}{gH}$ in equation (19) is omitted. If the Earth rotation rate is used, the difference between equation (5) and (19) will be negligible, but the process time for the test will drastically increase.

The implementation of NCTs has been a compiler option in the MPAS atmospheric dynamical core (Skamarock et al. 2012), which is fully compressible. Testing this option with the compressional Rossby waves, this study identified a flaw in its source code (the vertical NCT had been mistakenly divided by the grid-cell area in m²) and corrected it. For the simulation, the grid mesh comprises regular hexagons of which a pair of opposite sides lies in the zonal direction. The zonal grid spacing is 5 km, so 400 grid cells cover the domain width. The domain depth is equally divided into 64 grid boxes, so the vertical grid spacing is 198.77 m. All physical parameterization schemes and Rayleigh damping are switched off. The results suggest that the numerical solutions reasonably conform to the analytical solutions in this study; the contours of the results almost overlap those on Figure 1. In terms of the Euclidean norm of the zonal velocity field, Figure 3 depicts the percentages of the difference between the numerical and the analytical solutions to the analytical solution. This normalized difference decreases with $u_0$; at the end of one wave period (24 hours), 1.315% for $u_0 = 0.09$ m s⁻¹, 0.811% for $u_0 = 0.045$ m s⁻¹, and 0.625% for $u_0 = 0.0225$ m s⁻¹. For the zonal velocity field output every 3,600 s, see compilation of graphics in pdf format in the supplemental material, where the thick black and thin green contours denote analytical and numerical solutions. The difference is small and can be substantially explained by the zonal advection of zonal velocity. This conformation validates the recent correction of the implementation of NCTs in the MPAS atmospheric dynamical core.

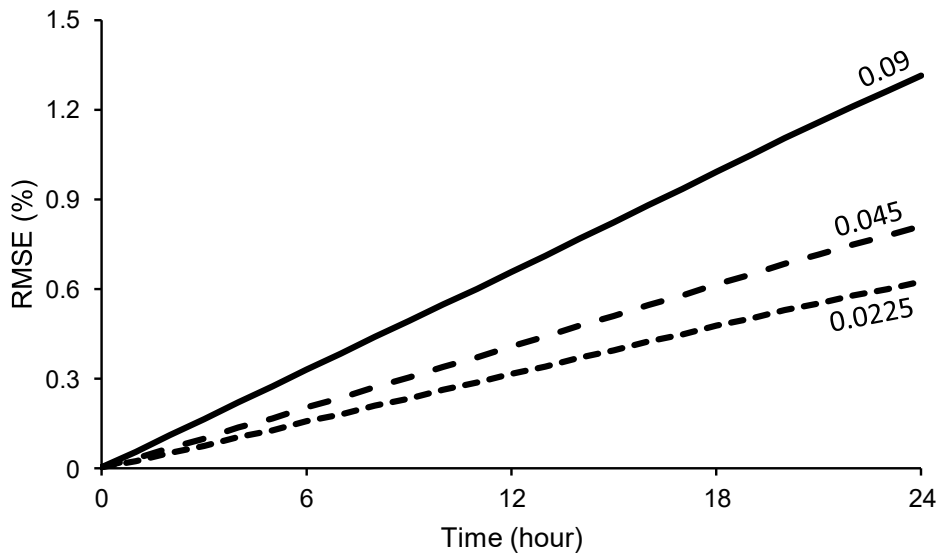

**Figure 3**. Temporal evolutions of the normalized root-mean-square errors of the numerical solutions of the compressional Rossby waves using the MPAS. The numbers by the curves denote the initial perturbation amplitude of zonal velocity in m s⁻¹.

## 6. Compressional beta effect in the MJO

To demonstrate how to analyze the compressional beta effect from reanalysis data, we take the MJO as an example, focusing on its zonal-vertical overturning circulation. The slow eastward phase speed of the compressional Rossby wave solutions motivates us to explore possible contributions of the compressional beta effect to the eastward propagation of the MJO, which is on the slowest end of the spectrum of Wheeler and Kiladis (1999) but 20 times on average faster than the compressional Rossby wave solutions. Although the model used to create reanalysis data does not include NCTs, the overturning circulations associated with the MJO in reanalysis data compare fairly well with those in observed data (Kiladis et al. 2005) and can be used to reconstruct the compressional beta effect. Accordingly, this study analyzes the MJO-filtered compressional beta effect and local meridional vorticity tendency reconstructed from ERA-Interim (Dee et al. 2011) reanalysis data from 1979 to 2018. The compressional beta effect is approximated from $-2\Omega \frac{w}{H}$ as equation (3) to $\frac{2\Omega}{p}\frac{Dp}{Dt}$, where $p$ denotes pressure, with the data in isobaric coordinates. The local meridional vorticity tendency is approximated with a central finite difference with a spacing of one day. An MJO index for every longitude is created by filtering tropical precipitation for an MJO band covering zonal wavenumber from 1 to 10 and time period from 30 days to 96 days. For the tropical precipitation, GPCP Version 1.3 One-Degree Daily Precipitation Data Set (Mesoscale Atmospheric Processes Branch and Earth System Science Interdisciplinary Center 2018) is averaged from 15°S to 15°N. Then, the compressional beta effect and the local meridional vorticity tendency are regressed upon the MJO index. The statistical significance is tested with two-tailed Student's t-test at 95% confidence level, where the equivalent degrees of freedom take autocorrelation of one-day lag into account.

Figure 4 depicts zonal vertical distributions at the equator of the results regressed upon the MJO-filtered precipitation at 90°E. The most prominent signal of the compressional beta effect is negative in the mid-upper troposphere in the MJO-active (convective) phase from 60°E to 135°E minimizing at 90°E. This negative compressional beta effect can be explained by upward motion associated with the MJO-active phase. The most prominent negative signal of the meridional vorticity tendency is collocated with the negative signal of the compressional beta effect. In terms of the ratio of the minimum values, the compressional beta effect contributes 10.8% of the meridional vorticity tendency. In other words, the east-up-west circulation in the west of the MJO-active phase is propagating toward the MJO-active phase partially owing to the compressional beta effect. The compressional beta effect is lacking in most of the current global atmospheric models because of the omission of NCTs, but the consequences of this lack may vary. For such models to yield an appropriate phase speed and amplitude of the MJO, they would need at least one of the other terms in equation (3) to overact, e.g., an overestimated west-east buoyancy gradient across the MJO-active phase. For the other terms to remain appropriate, the phase speed would be underestimated to maintain the amplitude, or the amplitude would decrease with time to maintain the phase speed. Another mechanism whereby NCTs can contribute to vorticity budgets is through tilting (Hayashi and Itoh 2012). We suspect that the tilting unlikely affects propagation for the following reasons. Adding only the tilting to Matsuno's (1966) model does not change the dispersion relations of the equatorial waves (Fruman 2009; Roundy and Janiga 2012). Adding both the tilting and the compressional beta effect to it yields additional eastward propagation (Section 4). Removing the tilting from this result by removing the $y$-dimension does not change the eastward propagation (Section 3).

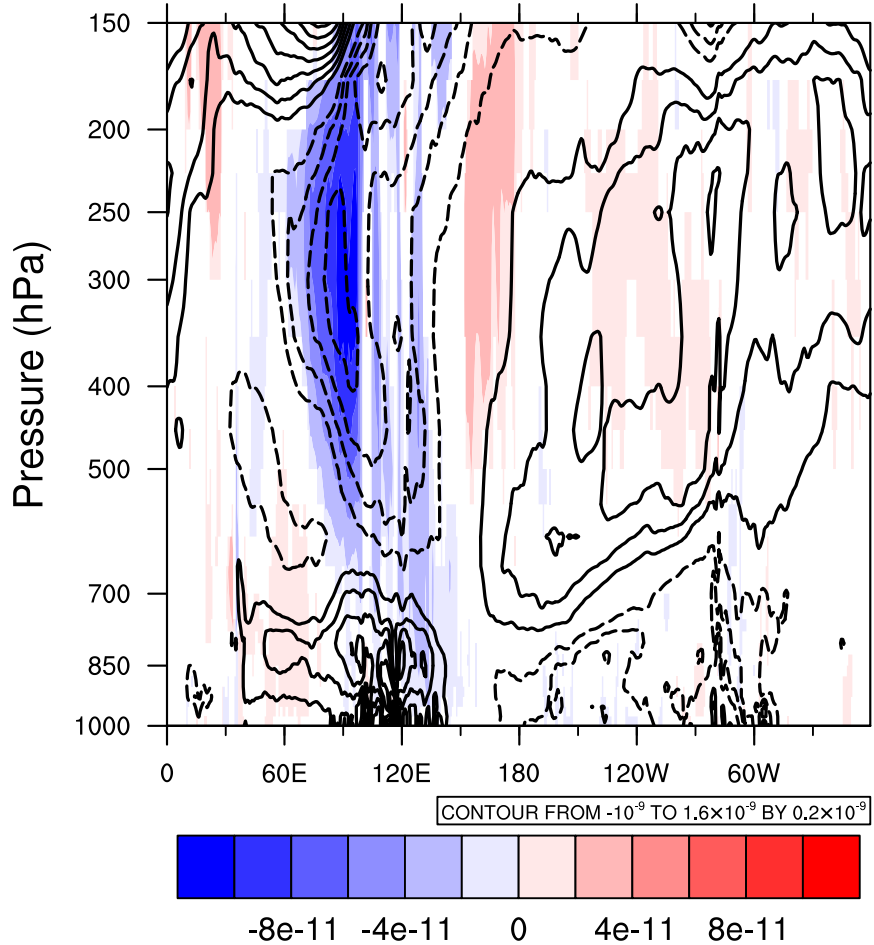

**Figure 4**. Zonal vertical distributions at the equator of the meridional vorticity tendency (contours, s$^{-2}$) and the compressional beta effect (shading, s$^{-2}$) regressed upon MJO-filtered tropical precipitation at 90°E. Significant at 95% confidence level, shown results are the prediction at one standard deviation of the filtered precipitation. The solid and dashed contours denote positive and negative values. The zero contour is omitted.

## 7. Summary and Discussion

This study corrects the derivation of the compressional Rossby wave solutions of Verhoeven and Stellmach (2014) by accounting for dynamical consistency and energy constraints. Compressional Rossby waves are meridional vorticity disturbances in the equatorial region that propagate eastward owing to the compressional beta effect. This effect is due to vertical density-weighted advection of the meridional planetary vorticity; the advected quantity is the meridional planetary vorticity divided by density, and multiplying density converts such an advection to local meridional relative vorticity tendency via compression or expansion. A signature of compressional Rossby waves is a low-pressure anomaly between easterly winds below and westerly winds above and a high with the opposite wind pattern. The compressional Rossby wave solutions can serve as a benchmark to validate the implementation of the nontraditional Coriolis terms (NCTs). With effectively neutral static stability and initial large-scale disturbances given a half vertical wavelength spanning the troposphere on Earth, compressional Rossby waves are expected to propagate eastward at a phase speed of 0.24 m s$^{-1}$. The phase speed increases with the planetary rotation rate and the vertical wavelength, and it also changes with the density scale height. This benchmark can be important because many model developers are restoring NCTs. We recently corrected the implementation of NCTs in the MPAS atmospheric dynamical core and validated the correction by simulating the compressional Rossby waves. This benchmarking test uses a generalized equatorial f-plane. Also, it uses fast planetary rotation rate to save process time. Nonetheless, it uses barotropic ideal gas to magnify the compressional beta effect without adding moist processes. The numerical solutions reasonably conform to the analytical solutions.

This study also derives a complete set of equatorially confined wave solutions from an anelastic equation set with the complete Coriolis terms, which include both the vertical and meridional planetary vorticity. The propagation mechanism can change with the effective static stability. In a strongly stable case in which the effective buoyancy frequency is larger than the meridional planetary vorticity, the dispersion relations appear like Matsuno's (1966), which is true for the canonical convectively coupled equatorial wave bands on Earth (e.g., Wheeler and Kiladis 1999). In the neutral case, in which buoyancy ceases, the compressional beta effect replaces buoyancy as the eastward-propagation mechanism, and westward-propagating modes that depend on buoyancy disappear. The complete set derived in this study remarkably differs from Matsuno's (1966) only if the meridional planetary vorticity is comparable or larger than the effective buoyancy frequency, which is consistent with Müller (1989).

As a demonstration of data analysis, the compressional beta effect and the meridional vorticity tendency are reconstructed using reanalysis data and regressed upon tropical precipitation data filtered for the MJO. In the mid-upper troposphere in the MJO-active phase, the compressional beta effect is prominently negative owing to the upward motion. In the same region, the meridional vorticity is decreasing with time. The compressional beta effect explains 10.8% of the decrease of the meridional vorticity in the MJO-active phase in terms of the ratio of the minimum values.

More consideration shall be given to theories about a dynamical continuum from the Kelvin waves to the MJO. Roundy (2020) showed that observed signals conforming to unforced Kelvin waves exist in the upper troposphere throughout the Kelvin-wave–MJO spectrum. Adames et al. (2019) included moisture variability into a zero-$\hat{v}$ wave framework, and the results suggest that the moisture dynamics becomes significant while the system is adjusted toward the MJO. The present study encourages a combination of both NCTs and moisture variability for future studies because NCTs are also potentially considerable for MJO propagation. Still, this combination may not combine the unforced wave framework and the forced-dissipative framework. Yet the MJO appears like unforced waves in the upper troposphere but like forced flow in the lower troposphere (Roundy 2012). This challenge is also left for future studies.

**Acknowledgments**

This work was funded by National Science Foundation (grants AGS1757342, AGS1358214, and AGS1128779). This paper originated as a course project of Hing Ong in ATM 523, Large Scale Dynamics of the Tropics, instructed by Paul Roundy. It became a chapter of Hing Ong's PhD dissertation, accepted by a committee composed of Paul Roundy, William Skamarock, Brian Rose, and Robert Fovell. We thank William Skamarock for discussion and technical support on the development of the benchmarking test. We thank Paul Roundy's (previously Hing Ong's) department for funding this paper and thank the other students in the class for discussion. Hing Ong was funded by Government Scholarship to Study Abroad, Ministry of Education, Taiwan. We thank Kai-Chih Tseng, Kevin Reed, and four anonymous reviewers for the useful comments.

Hing Ong thanks especially an anonymous student reviewer in the class. We thank ECMWF for granting access to ERA-Interim data via NCAR Research Data Archive.

**Data Availability Statement**

The source code generating analytical solutions for the compressional Rossby waves are available from https://github.com/HingOng/CompressionalRossbyWave. The source code of the MPAS and the mesh file for the test case can be obtained via https://github.com/MPAS-Dev/MPAS-Model and https://www2.mmm.ucar.edu/projects/mpas/test_cases/v7.0/mountain_wave.tar.gz. GPCP Version 1.3 and ERA-Interim data can be obtained via https://doi.org/10.5065/PV8B-HV76 and https://doi.org/10.5065/D6CR5RD9.